\documentclass[twoside,a4paper]{amsart}
\usepackage{graphicx}
\usepackage{xr-hyper}
\PassOptionsToPackage{pdfauthor={Vladimir V. Kisil},%
    pdftitle={},%
    pdfsubject={mathematics},%
    backref=page,%
  pdfkeywords={symmetries, quantum mechanics, classical mechanics,
    non-standard analysis, dual numbers, double numbers}
}{hyperref}
\IfFileExists{eulervm.sty}{\usepackage{eulervm}

\newcommand{\Aprime}{A\!'}}{\newcommand{\Aprime}{A'}}
\newcommand{\ppnum}[3]{\def\publname{\texttt{#2}, #3\\
    #1}}
\usepackage{amssymb}
\ppnum{%LEEDS--MATH--PURE--200?-??
}{arXiv: \href{http://arXiv.org/abs/1103.1120}{arXiv:1103.1120}}
{2011}

\PII{\relax}
\copyrightinfo{}{\relax}
\input{cyr.fd}

\newtheorem{thm}{Theorem}
\newtheorem{prop}[thm]{Proposition}

\newtheorem{principle}[thm]{Principle}

\theoremstyle{definition}

\newtheorem{example}[thm]{Example}

\theoremstyle{remark}
\newtheorem{rem}[thm]{Remark}
\newcommand{\AMSMSC}[2]{\subjclass[2000]{Primary #1; Secondary #2.}}

\usepackage[breaklinks=true,%
colorlinks=true,%
bookmarks=true,%
backref=page,%
pagebackref=true]{hyperref}

\usepackage[backrefs,msc-links
]{amsrefs}

\providecommand{\Sp}[1][n]{\ensuremath{\mathrm{Sp}(#1)}}
\providecommand{\SL}[1][2]{\ensuremath{\mathrm{SL}_{#1}(\Space{R}{})}}
\providecommand{\rmi}{\mathrm{i}}
\providecommand{\rmp}{\varepsilon}
\providecommand{\rmh}{\mathrm{j}}
\providecommand{\ladder}[2][]{L_{#1}^{\!#2}}

\usepackage[all]{xy}

\providecommand{\algebra}[1]{\ensuremath{\mathfrak{#1}}}
\providecommand{\myh}{h}
\providecommand{\myhbar}{\hslash}

\makeatletter
  \providecommand{\limlike}[1]{\mathop {\operator@font #1}}
  \providecommand{\loglike}[1]{\mathop {\operator@font #1}\nolimits}
\makeatother
\providecommand{\rmh}{\mathrm{j}}

\providecommand{\rmp}{\varepsilon}

\providecommand{\alli}{\iota}
\providecommand{\ladder}[2][]{L_{#1}^{\!#2}}
\providecommand{\uir}[3][0]{\ifcase #1{\rho^{#2}_{#3}}% 
\or {\breve{\rho}^{#2}_{#3}}%
\or {\tilde{\rho}^{#2}_{#3}}\fi}

\providecommand{\comment}[1]{}
\providecommand{\Space}[3][]{\ensuremath{\mathbb{#2}^{#3}_{#1}{}}}
  \providecommand{\FSpace}[3][]{\ensuremath{\ifx#2l \ell_{#3}^{#1}{}\else
  #2_{#3}^{#1}{}\fi}} 

\providecommand{\keywords}[1]{}
\providecommand{\urladdr}[1]{}

\usepackage{noweb}
\begin{document}
\title[EPAL3.2 Ladder Operators in Hypercomplex Mechanics]{Erlangen Programme at Large 3.2\\ 
  Ladder Operators in Hypercomplex Mechanics}

\author[Vladimir V. Kisil]%
{\href{http://www.maths.leeds.ac.uk/~kisilv/}{Vladimir V. Kisil}}
\thanks{On  leave from Odessa University.}

\address{%            
%Institute of Mathematics\\
%Economics and Mechanics\\
%Odessa State University\\
%ul. Petra Velikogo, 2\\
%Odessa-57, 270057, UKRAINE
School of Mathematics\\
University of Leeds\\
Leeds LS2\,9JT\\
UK
}

\email{\href{mailto:kisilv@maths.leeds.ac.uk}{kisilv@maths.leeds.ac.uk}}

\urladdr{\href{http://www.maths.leeds.ac.uk/~kisilv/}%
{http://www.maths.leeds.ac.uk/\~{}kisilv/}}

\dedicatory{Dedicated to the memory of Ian R. Porteous}

\begin{abstract}
  We revise the construction of creation/annihilation operators in
  quantum mechanics based on the representation theory of the
  Heisenberg and symplectic groups. Besides the standard harmonic
  oscillator (the elliptic case) we similarly treat the repulsive
  oscillator (hyperbolic case) and the free particle (the parabolic
  case).  The respective hypercomplex numbers turn out to be handy on this
  occasion. This provides a further illustration to the Similarity and
  Correspondence Principle.
\end{abstract}
\keywords{Heisenberg group, Kirillov's method of orbits, geometric
  quantisation, quantum mechanics, classical mechanics, Planck
  constant, dual numbers, double numbers, hypercomplex, jet spaces,
  hyperbolic mechanics, interference, Fock--Segal--Bargmann representation,
  Schr\"odinger representation, dynamics equation, harmonic and
  unharmonic oscillator, contextual probability, symplectic group,
  metaplectic representation, Shale--Weil representation} 
\AMSMSC{81R05}{81R15, 22E27, 22E70, 30G35, 43A65}
\maketitle
\tableofcontents
\section{Introduction}
\label{sec:introduction}
%  This is a report on the ongoing work on unification of quantum and
%  classical mechanical formalism, known as p-mechanics. 

Harmonic oscillators are treated in most
textbooks on quantum mechanics. This is efficiently done through
creation/annihilation (ladder) operators~\citelist{\cite{Gazeau09a}
  \cite{BoyerMiller74a}}. The underlying structure is the representation
theory of the the Heisenberg and symplectic
groups~\citelist{\cite{Lang85}*{\S~VI.2} \cite{MTaylor86}*{\S~8.2}
  \cite{Howe80b} \cite{Folland89}}. It is also known that quantum
mechanics and field theory can benefit from the introduction of Clifford
algebra-valued group representations~\citelist{\cite{Kisil93c}
  \cite{ConstalesFaustinoKrausshar11a} \cite{CnopsKisil97a}
  \cite{GuentherKuzhel10a}}. 

The dynamics of a harmonic oscillator generates the symplectic
transformation of the phase space of the elliptic type. The respective
parabolic and hyperbolic counterparts are also of
interest~\citelist{\cite{Wulfman10a}*{\S~3.8} \cite{ATorre08a}}. As we
will see, they are naturally connected with the respective hypercomplex
numbers.

To make this correspondence explicit we recall that the symplectic
group \(\Sp[2]\)~\cite{Folland89}*{\S~1.2} consists of
\(2\times 2\) matrices with real entries and the unit determinant. It
is isomorphic to the group \(\SL\)~\citelist{ \cite{Lang85}
  \cite{HoweTan92} \cite{Mazorchuk09a}} and provides linear
symplectomorphisms of the two-dimensional phase space. It has three
types of non-isomorphic one-dimensional subgroups represented by:
\begin{eqnarray}
  \label{eq:k-subgroup}
  K&=&\left\{ {\begin{pmatrix}
        \cos t &  \sin t\\
        -\sin t & \cos t
      \end{pmatrix}=   \exp \begin{pmatrix} 0& t\\-t&0
      \end{pmatrix}},\ t\in(-\pi,\pi]\right\},\\
  \label{eq:n-subgroup}
  N&=&\left\{   {\begin{pmatrix} 1&t \\0&1
      \end{pmatrix}=\exp \begin{pmatrix} 0 & t\\0&0
      \end{pmatrix},}\  t\in\Space{R}{}\right\},\\
  \label{eq:a-subgroup}
  A&=&\left\{  
    \begin{pmatrix} e^t & 0\\0&e^{-t}
    \end{pmatrix}=\exp \begin{pmatrix} t & 0\\0&-t
    \end{pmatrix},\  t\in\Space{R}{}\right\}.
\end{eqnarray}
We will refer to them as elliptic, parabolic and hyperbolic subgroups,
respectively. 

On the other hand, there are three non-isomorphic types of commutative,
associative two-dimensional algebras known as complex, dual and
double numbers~\citelist{\cite{Yaglom79}*{App.~C}
  \cite{LavrentShabat77}*{\S~5}}. They are represented by expressions
\(x+\alli y\), where \(\alli\) stands for one of the hypercomplex units
\(\rmi\), \(\rmp\) or \(\rmh\) with the properties:
\begin{displaymath}
  \rmi^2=-1, \qquad \rmp^2=0, \qquad \rmh^2=1.
\end{displaymath}
These units can also be labelled as elliptic, parabolic and hyperbolic.

In an earlier paper~\cite{Kisil10a},we considered representations of
the Heisenberg group which are induced by hypercomplex characters of
its centre. The elliptic case (complex numbers) describesthe
traditional framework of quantum mechanics, of course.

Double-valued representations, with the imaginary unit
\(\rmh^2=1\), are atural source of hyperbolic quantum mechanics
developed for a
while~\cites{Hudson66a,Hudson04a,Khrennikov03a,Khrennikov05a,Khrennikov08a}.
The representation acts on a Krein space with 
an indefinite inner product~\cite{AzizovIokhvidov71a}. This aroused
significant recent interest in connection with
\(\mathcal{PT}\)--symmetric quantum
mechanics~\cite{GuentherKuzhel10a}. However, our approach is different
from the classical treatment of Krein spaces: we use the hyperbolic
unit \(\rmh\) and build the hyperbolic analytic function theory on its
own basis~\cites{Kisil97c,Kisil11c}. In the traditional approach, the
indefinite metric is mapped to a definite inner product through an
auxiliary operators.

The representation with values in dual numbers provides a convenient
description of the classical mechanics. To this end we do not take any
sort of semiclassical limit, rather the nilpotency of the imaginary
unit (\(\rmp^2=0\)) performs the task. This removes the vicious
necessity to consider the Planck \emph{constant} tending to zero.
Mixing this with complex numbers we get a convenient tool for
modelling the interaction between quantum and classical
systems~\cites{Kisil05c,Kisil09b}.

Our construction~\cite{Kisil10a} provides three different types of
dynamics and also generates the respective rules for addition of
probabilities. In this paper we analyse the three types of dynamics
produced by transformations~(\ref{eq:k-subgroup}--\ref{eq:a-subgroup})
from the symplectic group \(\Sp[2]\) by means of ladder
operators. As a result we obtain further illustrations to the following:
\begin{principle}[Similarity and Correspondence]
  \cite{Kisil09c}*{Principle~\ref{W-pr:simil-corr-principle}}
  \label{pr:similarity-correspondence}
  \begin{enumerate}
  \item Subgroups \(K\), \(N\) and \(A\) play a similar r\^ole in the
    structure of the group \(\Sp[2]\) and its representations.
  \item The subgroups shall be swapped simultaneously with the
    respective replacement of hypercomplex unit \(\alli\).
  \end{enumerate}
\end{principle}
Here the two parts are interrelated: without a swap of imaginary units
there can be no similarity between different subgroups.

In this paper we work with the simplest case of a particle with only one
degree of freedom. Higher dimensions and the respective group of
symplectomorphisms \(\Sp[2n]\) may require consideration of Clifford
algebras~\cite{Porteous95}.

\section{Heisenberg Group and Its Automorphisms}
\label{sec:prel-heis-group}

Let \((s,x,y)\), where \(s\), \(x\), \(y\in \Space{R}{}\), be
an element of the one-dimensional Heisenberg group
\(\Space{H}{1}\)~\cites{Folland89,Howe80b}. Consideration of the
general case of \(\Space{H}{n}\) will be similar, but is beyond  the
scope of present paper. The group law on
\(\Space{H}{1}\) is given as follows:
\begin{equation}
  \label{eq:H-n-group-law}
  \textstyle
  (s,x,y)\cdot(s',x',y')=(s+s'+\frac{1}{2}\omega(x,y;x',y'),x+x',y+y'), 
\end{equation} 
where the non-commutativity is due to \(\omega\)---the
\emph{symplectic form} on \(\Space{R}{2n}\)~\cite{Arnold91}*{\S~37}:
\begin{equation}
  \label{eq:symplectic-form}
  \omega(x,y;x',y')=xy'-x'y.
\end{equation}
The Heisenberg group is a non-commutative Lie
group. % with the centre
%\begin{displaymath}
%  Z=\{(s,0,0)\in \Space{H}{1}, \ s \in \Space{R}{}\}.
%\end{displaymath}
The left shifts
\begin{equation}
  \label{eq:left-right-regular}
  \Lambda(g): f(g') \mapsto f(g^{-1}g')  
\end{equation}
act as a representation of \(\Space{H}{1}\) on a certain linear space
of functions. For example, an action on \(\FSpace{L}{2}(\Space{H}{},dg)\) with
respect to the Haar measure \(dg=ds\,dx\,dy\) is the \emph{left regular}
representation, which is unitary.

The Lie algebra \(\algebra{h}^n\) of \(\Space{H}{1}\) is spanned by
left-(right-)invariant vector fields
\begin{equation}
\textstyle  S^{l(r)}=\pm{\partial_s}, \quad
  X^{l(r)}=\pm\partial_{ x}-\frac{1}{2}y{\partial_s},  \quad
 Y^{l(r)}=\pm\partial_{y}+\frac{1}{2}x{\partial_s}
  \label{eq:h-lie-algebra}
\end{equation}
on \(\Space{H}{1}\) with the Heisenberg \emph{commutator relation} 
\begin{equation}
  \label{eq:heisenberg-comm}
  [X^{l(r)},Y^{l(r)}]=S^{l(r)} 
\end{equation}
and  all other commutators vanishing. We will sometime omit the superscript \(l\)
for left-invariant field. 

The group of outer automorphisms of \(\Space{H}{1}\), which trivially
acts on the centre of \(\Space{H}{1}\), is the symplectic group
\(\Sp[2]\) defined in the precious section. It is the group of
symmetries of the symplectic form
\(\omega\)~\citelist{\cite{Folland89}*{Thm.~1.22}
  \cite{Howe80a}*{p.~830}}. The symplectic group is isomorphic to
\(\SL\)~\citelist{\cite{Lang85} \cite{MTaylor86}*{Ch.~8}}. The
explicit action of \(\Sp[2]\) on the Heisenberg group is:
\begin{equation}
  \label{eq:sympl-auto}
  g: h=(s,x,y)\mapsto g(h)=(s,x',y'), 
\end{equation}
where 
\begin{displaymath}
  g=\begin{pmatrix}
    a&b\\
    c&d
  \end{pmatrix}\in\Sp[2], \quad\text{ and }\quad
  \begin{pmatrix}
    x'\\y'
  \end{pmatrix}
  =\begin{pmatrix}
    a&b\\
    c&d
  \end{pmatrix}
  \begin{pmatrix}
    x\\y
  \end{pmatrix}.
\end{displaymath}
The Shale--Weil theorem~\citelist{\cite{Folland89}*{\S~4.2}
  \cite{Howe80a}*{p.~830}} states that any representation
\(\uir{}{\myhbar}\) of the Heisenberg groups generates a unitary
\emph{oscillator} (or \emph{metaplectic}) representation
\(\uir{\text{SW}}{\myhbar}\) of the \(\widetilde{\mathrm{Sp}}(2)\),
the two-fold cover of the symplectic group~\cite{Folland89}*{Thm.~4.58}.

We can consider the semidirect product
\(G=\Space{H}{1}\rtimes\widetilde{\mathrm{Sp}}(2)\) with the standard group law:
\begin{displaymath}
  (h,g)*(h',g')=(h*g(h'),g*g'), \qquad \text{where } 
  h,h'\in\Space{H}{1}, \quad g,g'\in\widetilde{\mathrm{Sp}}(2),
\end{displaymath}
and the stars denote the respective group operations while the action
\(g(h')\) is defined as the composition of the projection map
\(\widetilde{\mathrm{Sp}}(2)\rightarrow {\mathrm{Sp}}(2)\) and the
action~\eqref{eq:sympl-auto}. This group is sometimes called the
Schr\"odinger group, and it is known as the maximal kinematical
invariance group of both the free Schr\"odinger equation and the
quantum harmonic oscillator~\cite{Niederer73a}. This group is of
interest not only in quantum mechanics but also in
optics~\cites{ATorre10a,ATorre08a}.

Consider the Lie algebra \(\algebra{sp}_2\) of the group
\(\Sp[2]\). Pick up the following basis in
\(\algebra{sp}_2\)~\cite{MTaylor86}*{\S~8.1}:
\begin{displaymath}
  A= \frac{1}{2}
  \begin{pmatrix}
    -1&0\\0&1
  \end{pmatrix},\quad 
  B= \frac{1}{2} \
  \begin{pmatrix}
    0&1\\1&0
  \end{pmatrix}, \quad 
  Z=
  \begin{pmatrix}
    0&1\\-1&0
  \end{pmatrix}.
\end{displaymath}
The commutation relations between the elements are:
\begin{equation}
  \label{eq:sl2-commutator}
  [Z,A]=2B, \qquad [Z,B]=-2A, \qquad [A,B]=\textstyle- \frac{1}{2} Z.
\end{equation} 
Vectors \(Z\), \(B+Z/2\) and \(-A\) are generators of the
one-parameter subgroups \(K\), \(N\) and \(A\)
(\ref{eq:k-subgroup}--\ref{eq:a-subgroup}) respectively.

\begin{figure}[htbp]
  \centering
  \includegraphics[scale=.8]{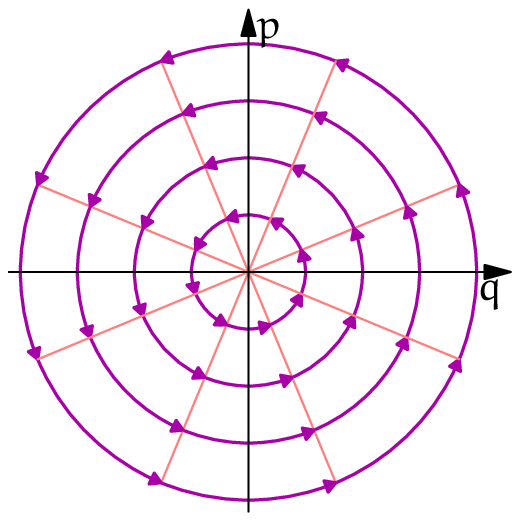}\hfill
  \includegraphics[scale=.8]{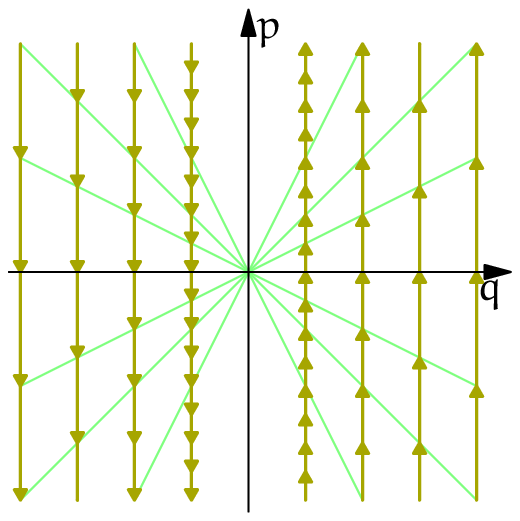}\hfill
  \includegraphics[scale=.8]{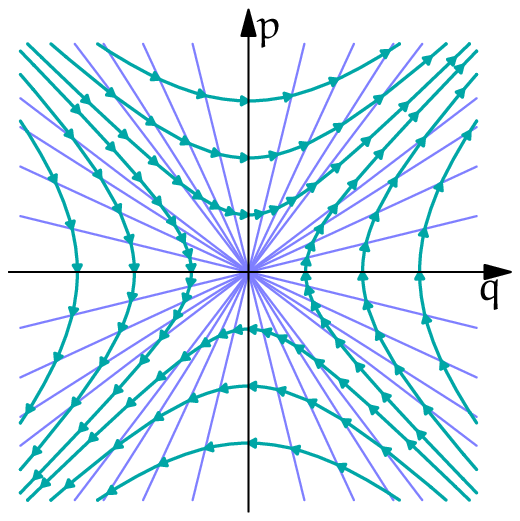}
  \caption{Three types (elliptic, parabolic and hyperbolic) of linear
    symplectic transformations on the plane}
  \label{fig:eph-symplect}
\end{figure}

Furthermore, we can consider the basis \(\{S, X, Y, A, B, Z\}\) of the
Lie algebra \(\algebra{g}\) of the Lie group
\(G=\Space{H}{1}\rtimes\widetilde{\mathrm{Sp}}(2)\). All non-zero
commutators besides those already listed in~\eqref{eq:heisenberg-comm}
and~\eqref{eq:sl2-commutator} are: 
\begin{align}
  \label{eq:cross-comm}
  [A,X]&=\textstyle\frac{1}{2}X,&
  [B,X]&=\textstyle-\frac{1}{2}Y,&
  [Z,X]&=Y;\\
  \label{eq:cross-comm1}
  [A,Y]&=\textstyle-\frac{1}{2}Y,&
  [B,Y]&=\textstyle-\frac{1}{2}X,&
  [Z,Y]&=-X.
\end{align}
The Shale--Weil theorem allows us to expand any representation
\(\uir{}{\myhbar}\) of the Heisenberg group to the representation
\(\uir[2]{}{\myhbar}=\uir{}{\myhbar}\oplus\uir{\text{SW}}{\myhbar}\) of
group \(G\).
\begin{example}
  Let \(\uir{}{\myhbar}\) be the Schr\"odinger
  representation~\citelist{\cite{Folland89}*{\S~1.3}
  } of \(\Space{H}{1}\) in
  \(\FSpace{L}{2}(\Space{R}{})\), that is \cite{Kisil10a}*{~\eqref{E-eq:schroedinger-rep-conf}}:
  \begin{equation}
    \label{eq:schroedinger-rep-conf}
    [\uir{}{\chi}(s,x,y) f\,](q)=e^{2\pi\rmi\myhbar (s-xy/2)
      +2\pi\rmi x q}\,f(q-\myhbar y).  
  \end{equation}
  Thus the action of the derived representation on the Lie algebra
  \(\algebra{h}_1\) is:
  \begin{equation}
    \label{eq:schroedinger-rep-conf-der}
    \uir{}{\myhbar}(X)=2\pi\rmi q,\qquad \uir{}{\myhbar}(Y)=-\myhbar \frac{d}{dq},
    \qquad
    \uir{}{\myhbar}(S)=2\pi\rmi\myhbar I.
  \end{equation}
  Then the associated Shale--Weil representation of \(\Sp[2]\) in
  \(\FSpace{L}{2}(\Space{R}{})\) has the
  derived action,  cf.~\citelist{\cite{ATorre08a}*{(2.2)} \cite{Folland89}*{\S~4.3}}:
  \begin{equation}
    \label{eq:shale-weil-der}
    \uir{\text{SW}}{\myhbar}(A) =-\frac{q}{2}\frac{d}{dq}-\frac{1}{4},\quad
    \uir{\text{SW}}{\myhbar}(B)=-\frac{\myhbar\rmi}{8\pi}\frac{d^2}{dq^2}-\frac{\pi\rmi q^2}{2\myhbar},\quad
    \uir{\text{SW}}{\myhbar}(Z)=\frac{\myhbar\rmi}{4\pi}\frac{d^2}{dq^2}-\frac{\pi\rmi q^2}{\myhbar}.
  \end{equation}
  We can verify commutators~\eqref{eq:heisenberg-comm} and
  (\ref{eq:sl2-commutator}--\ref{eq:cross-comm1}) for
  operators~(\ref{eq:schroedinger-rep-conf-der}--\ref{eq:shale-weil-der}).
  It is also obvious that in this representation the following
  algebraic relations hold:
  \begin{eqnarray}
    \label{eq:quadratic-A}
    \qquad\uir{\text{SW}}{\myhbar}(A) &=&
    \frac{\rmi}{4\pi\myhbar}(\uir{}{\myhbar}(X)\uir{}{\myhbar}(Y)-{\textstyle\frac{1}{2}}\uir{}{\myhbar}(S))
    =\frac{\rmi}{8\pi\myhbar}(\uir{}{\myhbar}(X)\uir{}{\myhbar}(Y)+\uir{}{\myhbar}(Y)\uir{}{\myhbar}(X) ),\\ 
    \label{eq:quadratic-B}
    \uir{\text{SW}}{\myhbar}(B) &=&
    \frac{\rmi}{8\pi\myhbar}(\uir{}{\myhbar}(X)^2-\uir{}{\myhbar}(Y)^2), \\
    \label{eq:quadratic-Z}
    \uir{\text{SW}}{\myhbar}(Z)
    &=&\frac{\rmi}{4\pi\myhbar}(\uir{}{\myhbar}(X)^2+\uir{}{\myhbar}(Y)^2). 
  \end{eqnarray}
  Thus it is common in quantum optics to name \(\algebra{g}\) as a Lie
  algebra with  quadratic generators, see~\cite{Gazeau09a}*{\S~2.2.4}.
\end{example}
Note that \(\uir{\text{SW}}{\myhbar}(Z)\) is the Hamiltonian of the
harmonic oscillator (up to a factor). Then we can consider
\(\uir{\text{SW}}{\myhbar}(B)\) as the Hamiltonian of a repulsive
(hyperbolic) oscillator. The operator
\(\uir{\text{SW}}{\myhbar}(B-Z/2)=\frac{\myhbar\rmi}{4\pi}\frac{d^2}{dq^2}\)
is the parabolic analog. A graphical representation of all three
transformations is given in Fig.~\ref{fig:eph-symplect}, and a further
discussion of these Hamiltonians can be found
in~\cite{Wulfman10a}*{\S~3.8}. 
An important observation, which is often missed, is that the
three linear symplectic transformations are unitary rotations in the
corresponding hypercomplex algebra. This means, that the
symplectomorphisms generated by operators \(Z\), \(B-Z/2\), \(B\)
within time \(t\) coincide with the multiplication of hypercomplex
number \(q+\alli p\) by \(e^{\alli t}\)~\cite{Kisil09c}*{\S~3}, which
is just another illustration of the Similarity and Correspondence
Principle~\ref{pr:similarity-correspondence}.
\begin{example}
  There are many advantages of considering representations of the
  Heisenberg group on the phase
  space~\citelist{\cite{Howe80b}*{\S~1.7} \cite{Folland89}*{\S~1.6}
    \cite{deGosson08a}}. A convenient expression for
  Fock--Segal--Bargmann (FSB) representation on the phase space
  is~\cite{Kisil10a}*{\eqref{E-eq:stone-inf}}:
  \begin{equation}
    \label{eq:stone-inf}
    \textstyle
    [\uir{}{F}(s,x,y) f] (q,p)=
    e^{-2\pi\rmi(\myhbar s+qx+py)}
    f \left(q-\frac{\myhbar}{2} y, p+\frac{\myhbar}{2} x\right).
  \end{equation}
  Then the derived representation of \(\algebra{h}_1\) is:
  \begin{equation}
    \label{eq:fock-rep-conf-der-par1}
    \textstyle
    \uir{}{F}(X)=-2\pi\rmi q+\frac{\myhbar}{2}\partial_{p},\qquad
    \uir{}{F}(Y)=-2\pi\rmi p-\frac{\myhbar}{2}\partial_{q},
    \qquad
    \uir{}{F}(S)=-2\pi\rmi\myhbar I.
  \end{equation}
  This produces the derived form of the Shale--Weil representation:
  \begin{equation}
    \label{eq:shale-weil-der-ell}
    \textstyle
    \uir{\text{SW}}{F}(A) =\frac{1}{2}\left(q\partial_{q}-p\partial_{p}\right),\quad
    \uir{\text{SW}}{F}(B)=-\frac{1}{2}\left(p\partial_{q}+q\partial_{p}\right),\quad
    \uir{\text{SW}}{F}(Z)=p\partial_{q}-q\partial_{p}.
  \end{equation}
  Note that this representation does not contain the parameter
  \(\myhbar\), unlike the equivalent
  representation~\eqref{eq:shale-weil-der}. Thus the FSB model
  explicitly shows the equivalence of \(\uir{\text{SW}}{\myhbar_1}\) and
  \(\uir{\text{SW}}{\myhbar_2}\) if \(\myhbar_1
  \myhbar_2>0\)~\cite{Folland89}*{Thm.~4.57}.  

  As we will also see below, the FSB-type representations in
  hypercomplex numbers produce almost the same Shale--Weil
  representations.
\end{example}

\section{Ladder Operators in Quantum Mechanics}
\label{sec:ladder-operators}

Let \(\uir{}{}\) be a representation of the group
\(G=\Space{H}{1}\rtimes\widetilde{\mathrm{Sp}}(2)\) in a space \(V\).
Consider the derived representation of the Lie algebra
\(\algebra{g}\)~\cite{Lang85}*{\S~VI.1} and denote
\(\tilde{X}=\uir{}{}(X)\) for \(X\in\algebra{g}\). To see the
structure of the representation \(\uir{}{}\) we can decompose the
space \(V\) into eigenspaces of the operator \(\tilde{X}\) for some
\(X\in \algebra{g}\). The canonical example is the Taylor series in
complex analysis.

We are going to consider three cases corresponding to three
non-isomorphic subgroups~(\ref{eq:k-subgroup}--\ref{eq:a-subgroup}) of
\(\Sp[2]\) starting from the compact case. Let \(H=Z\) be a
generator of the compact subgroup \(K\).  Corresponding
symplectomorphisms~\eqref{eq:sympl-auto} of the phase space are given
by orthogonal rotations with matrices \(\begin{pmatrix} \cos t & \sin
  t\\ -sin t& \cos t
\end{pmatrix}\). The Shale--Weil
representation~\eqref{eq:shale-weil-der} coincides with the Hamiltonian of
the harmonic oscillator.

Since this is a double cover of a compact group, the corresponding
eigenspaces \(\tilde{Z} v_k=\rmi k v_k\) are parametrised by a
half-integer \(k\in\Space{Z}{}/2\). Explicitly for a half-integer \(k\):
\begin{equation}
  \label{eq:hermit-poly}
  v_k(q)=H_{k+\frac{1}{2}}\left(\sqrt{\frac{2\pi}{\myhbar}}q\right) e^{-\frac{\pi}{\myhbar}q^2},
\end{equation}
where \(H_k\) is the Hermite polynomial%
%The corresponding eigenvectors are Weber--Hermit functions
~\citelist{\cite{Folland89}*{\S~1.7}  \cite{ErdelyiMagnusII}*{8.2(9)}}.  

From the point of view of quantum mechanics and the
representation theory (which may be the same), it is beneficial to
introduce the ladder operators \(\ladder{\pm}\), known as
\emph{creation/annihilation} in quantum
mechanics~\cite{Folland89}*{p.~49} or \emph{raising/lowering} in
representation theory~\citelist{\cite{Lang85}*{\S~VI.2}
  \cite{MTaylor86}*{\S~8.2} \cite{BoyerMiller74a}}. They are defined by the following
commutation relations:
\begin{equation}
  \label{eq:raising-lowering}
  [\tilde{Z},\ladder{\pm}]=\lambda_\pm \ladder{\pm}. 
\end{equation}
In other words, \(\ladder{\pm}\) are eigenvectors for operators
\(\loglike{ad}Z\) of the adjoint representation of
\(\algebra{g}\)~\cite{Lang85}*{\S~VI.2}. 
\begin{rem}
  The existence of such ladder operators follows from the general
  properties of Lie algebras if the Hamiltonian belongs to a Cartan
  subalgebra. This is the case for vectors \(Z\) and \(B\), which are
  the only two non-isomorphic types of Cartan subalgebras in
  \(\algebra{sl}_2\). However, the third case considered in this paper,
  the parabolic vector \(B+Z/2\), does not belong to a Cartan
  subalgebra, yet a sort of ladder operators is still possible with
  dual number coefficients. Moreover, for the hyperbolic vector \(B\),
  besides the standard ladder operators an additional pair with double
  number coefficients will also be described.
\end{rem}

From the commutators~\eqref{eq:raising-lowering} we deduce that if
\(v_k\) is an eigenvector of \(\tilde{Z}\) then \(\ladder{+} v_k\) is
an eigenvector as well:
\begin{eqnarray}
  \tilde{Z}(\ladder{+} v_k)&=&(\ladder{+}\tilde{Z}+\lambda_+\ladder{+})v_k=\ladder{+}(\tilde{Z}v_k)+\lambda_+\ladder{+}v_k
  =\rmi k \ladder{+}v_k+\lambda_+\ladder{+}v_k\nonumber \\
  &=&(\rmi k+\lambda_+)\ladder{+}v_k.
  \label{eq:ladder-action}
\end{eqnarray}
% Assuming \(\ladder{+}=aA+bB+cZ\) we
Thus the action of ladder operators on the respective eigenspaces
\(V_k\) can be visualised by the diagram:
\begin{equation}
  \label{eq:single-chain}
  \xymatrix@1{
    \ldots\, \ar@<.4ex>[r]^{\ladder{+}} &
    \,V_{\rmi k-\lambda}\,  \ar@<.4ex>[l]^{\ladder{-}}\ar@<.4ex>[r]^{\ladder{+}} &
    \,V_{\rmi k}\, \ar@<.4ex>[l]^{\ladder{-}} \ar@<.4ex>[r]^{\ladder{+}} &
    \,V_{\rmi k+ \lambda}\,\ar@<.4ex>[l]^{\ladder{-}}  \ar@<.4ex>[r]^{\ladder{+}}
    &
    \,\ldots\ar@<.4ex>[l]^{\ladder{-}}}
\end{equation}
There are two ways to search for ladder operators: in
(complexified) Lie algebras \(\algebra{h}_1\) and \(\algebra{sp}_2\).
We will consider them in a sequence.  

\subsection{Ladder Operators from the Heisenberg Group}
\label{sec:heis-group-oper}
Assuming \(\ladder{+}=a\tilde{X}+b\tilde{Y}\) we obtain from the
relations~(\ref{eq:cross-comm}--\ref{eq:cross-comm1})
and~\eqref{eq:raising-lowering} the linear equations with unknown
\(a\) and \(b\):
\begin{displaymath}
  a=\lambda_+ b, \qquad -b=\lambda_+ a.
\end{displaymath}
The equations have a solution if and only if \(\lambda_+^2+1=0\), and
the raising/lowering operators are \(\ladder{\pm}=
\tilde{X}\mp\rmi\tilde{Y}\). 
\begin{rem}
  Here we have an interesting asymmetric response: due to the
  structure of the semidirect product
  \(\Space{H}{1}\rtimes\widetilde{\mathrm{Sp}}(2)\) it is the
  symplectic group which acts on \(\Space{H}{1}\), not vise versa.
  However, the Heisenberg group has a weak action in the opposite
  direction: it shifts eigenfunctions of \(\Sp[2]\).
\end{rem}

In the Schr\"odinger
representation~\eqref{eq:schroedinger-rep-conf-der} the ladder
operators are
\begin{equation}
  \label{eq:ell-ladder-heisen-rep}
  \uir{}{\myhbar}(\ladder{\pm})= 2\pi\rmi q\pm\rmi\myhbar \frac{d}{dq}.
\end{equation}
The standard treatment of the harmonic oscillator in quantum mechanics,
which can be found in many textbooks, e.g.~\citelist{
  \cite{Folland89}*{\S~1.7} \cite{Gazeau09a}*{\S~2.2.3}}, 
is as follows. The vector  \(v_{-1/2}(q)=e^{-\pi q^2/\myhbar}\) is an
eigenvector of \(\tilde{Z}\) with the eigenvalue
\(-\frac{\rmi}{2}\). In addition \(v_{-1/2}\) is annihilated by
\(\ladder{+}\). Thus the chain~\eqref{eq:single-chain} terminates to
the right and the complete set of eigenvectors of the harmonic
oscillator Hamiltonian is presented by \((\ladder{-})^k v_{-1/2}\)
with \(k=0, 1, 2, \ldots\).

We can make a wavelet transform generated by the Heisenberg group with
the mother wavelet \(v_{-1/2}\), and the image will be the
Fock--Segal--Bargmann (FSB) space \citelist{\cite{Howe80b}
  \cite{Folland89}*{\S~1.6}}. Since \(v_{-1/2}\) is the null solution
of \(\ladder{+}=\tilde{X}-\rmi \tilde{Y}\), then by the general
result~\cite{Kisil10c}*{Cor.~\ref{C-co:cauchy-riemann}} the image of
the wavelet transform will be null-solutions of the corresponding linear
combination of the Lie derivatives~\eqref{eq:h-lie-algebra}:
\begin{equation}
  \label{eq:CR-Bargmann}
  D=\overline{X^{r} -\rmi  Y^{r}}=(\partial_{ x} +\rmi\partial_{y})-\pi\myhbar(x-\rmi
y),
\end{equation}
which turns out to be the Cauchy--Riemann equation on a weighted
FSB-type space. 

\subsection{Symplectic Ladder Operators}
\label{sec:sympl-ladd-oper}
We can also look for ladder operators within the Lie algebra
\(\algebra{sp}_2\), see~\cite{Kisil09c}*{\S~8}.
Assuming \(\ladder[2]{+}=a\tilde{A}+b\tilde{B}+c\tilde{Z}\) from the
relations~\eqref{eq:sl2-commutator} and defining
condition~\eqref{eq:raising-lowering} we obtain the linear equations
with unknown \(a\), \(b\) and \(c\): 
\begin{displaymath}
  c=0, \qquad 2a=\lambda_+ b, \qquad -2b=\lambda_+ a.
\end{displaymath}
The equations have a solution if and only if \(\lambda_+^2+4=0\), and
the raising/lowering operators are \(\ladder[2]{\pm}=\pm\rmi
\tilde{A}+\tilde{B}\). In the Shale--Weil
representation~\eqref{eq:shale-weil-der} they turn out to be:
\begin{equation}
  \label{eq:ell-ladder-symplect}
  \ladder[2]{\pm}=\pm\rmi\left(\frac{q}{2}\frac{d}{dq}+\frac{1}{4}\right)-\frac{\myhbar\rmi}{8\pi}\frac{d^2}{dq^2}-\frac{\pi\rmi q^2}{2\myhbar}=-\frac{\rmi}{8\pi\myhbar}\left(\mp2\pi q+\myhbar\frac{d}{dq}\right)^2.
\end{equation}
Since this time \(\lambda_+=2\rmi\) the ladder operators
\(\ladder[2]{\pm}\) produce a shift on the
diagram~\eqref{eq:single-chain} twice bigger than the operators
\(\ladder{\pm}\) from the Heisenberg group. After all, this is not
surprising since from the explicit
representations~\eqref{eq:ell-ladder-heisen-rep} and~\eqref{eq:ell-ladder-symplect} we get:
\begin{displaymath}
  \ladder[2]{\pm}=-\frac{\rmi}{8\pi\myhbar}(\ladder{\pm})^2.
\end{displaymath}

\section{Ladder Operators for the Hyperbolic Subgroup}
\label{sec:hiperbolic-subgroup}

Consider the case of the Hamiltonian \(H=2B\), which is a repulsive
(hyperbolic) harmonic oscillator~\cite{Wulfman10a}*{\S~3.8}. The
corresponding one-dimensional subgroup of symplectomorphisms produces
hyperbolic rotations of the phase space. The eigenvectors \(v_\mu\) of
the operator
\begin{displaymath}
  \uir{\text{SW}}{\myhbar}(2B)v_\nu
  =-\rmi\left(\frac{\myhbar}{4\pi}\frac{d^2}{dq^2}+\frac{\pi q^2}{\myhbar}\right)v_\nu
  =\rmi\nu v_\nu, 
\end{displaymath}
are Weber--Hermite (or parabolic cylinder) functions
\(v_{\nu}=D_{\nu-\frac{1}{2}}\left(\pm2e^{\rmi \frac{\pi}{4}}\sqrt{\frac{\pi}{\myhbar}} q\right)\),
see~\citelist{\cite{ErdelyiMagnusII}*{\S~8.2}
  \cite{SrivastavaTuanYakubovich00a}} for fundamentals of
Weber--Hermite functions and~\cite{ATorre08a} for further
illustrations and applications in optics.

The corresponding one-parameter group is not compact and the
eigenvalues of the operator \(2\tilde{B}\) are not restricted by any
integrality condition, but the raising/lowering operators are
still important~\citelist{\cite{HoweTan92}*{\S~II.1}
  \cite{Mazorchuk09a}*{\S~1.1}}. We again seek solutions in two
subalgebras \(\algebra{h}_1\) and \(\algebra{sp}_2\) separately.
However, the additional options will be provided by a choice of the
number system: either complex or double.

\subsection{Complex Ladder Operators}
\label{sec:compl-ladd-oper}

Assuming
\(\ladder[h]{+}=a\tilde{X}+b\tilde{Y}\) from the
commutators~(\ref{eq:cross-comm}--\ref{eq:cross-comm1}), we obtain
the linear equations:
\begin{equation}
  \label{eq:hyp-ladder-compatib}
  -a=\lambda_+ b, \qquad -b=\lambda_+ a.
\end{equation}
The equations have a solution if and only if \(\lambda_+^2-1=0\).
Taking the real roots \(\lambda=\pm1\) we obtain that the raising/lowering
operators are \(\ladder[h]{\pm}=\tilde{X}\mp\tilde{Y}\).  In the
Schr\"odinger representation~\eqref{eq:schroedinger-rep-conf-der} the
ladder operators are
\begin{equation}
  \label{eq:ell-ladder-heisen-rep1}
  \ladder[h]{\pm}= 2\pi\rmi q\pm \myhbar \frac{d}{dq}.
\end{equation}
The null solutions \(v_{\pm\frac{1}{2}}(q)=e^{\pm\frac{\pi\rmi}{\myhbar}
  q^2}\) to operators \(\uir{}{\myhbar}(\ladder{\pm})\) are also
eigenvectors of the Hamiltonian \(\uir{\text{SW}}{\myhbar}(2B)\) with the
eigenvalue \(\pm\frac{1}{2}\).  However the important distinction from
the elliptic case is, that they are not square-integrable on the real line
anymore.

We can also look for ladder operators within the \(\algebra{sp}_2\),
that is in the form \(\ladder[2h]{+}=a\tilde{A}+b\tilde{B}+c\tilde{Z}\)
for the commutator \([2\tilde{B},\ladder[h]{+}]=\lambda
\ladder[h]{+}\). We will get the system:
\begin{displaymath}
  4c=\lambda a,\qquad
  b=0,\qquad
  a=\lambda c.
\end{displaymath}
A solution again exists if and only if \(\lambda^2=4\). Within complex
numbers we get only the values \(\lambda=\pm 2\) with the ladder
operators \(\ladder[2h]{\pm}=\pm2\tilde{A}+\tilde{Z}/2\),
see~\citelist{\cite{HoweTan92}*{\S~II.1}
  \cite{Mazorchuk09a}*{\S~1.1}}. Each indecomposable \(\algebra{h}_1\)-
or \(\algebra{sp}_2\)-module is formed by a one-dimensional chain of
eigenvalues with a transitive action of ladder operators \(\ladder[h]{\pm}\)
or \(\ladder[2h]{\pm}\) respectively. And we again have a
quadratic relation between the ladder operators:
\begin{displaymath}
  \ladder[2h]{\pm}=\frac{\rmi}{4\pi\myhbar}(\ladder[h]{\pm})^2.
\end{displaymath}

\subsection{Double Ladder Operators}
\label{sec:double-ladd-oper}

There are extra possibilities in in the context of hyperbolic quantum
mechanics~\citelist{\cite{Khrennikov03a} \cite{Khrennikov05a}
  \cite{Khrennikov08a}}.  Here we use the representation of
\(\Space{H}{1}\) induced by a hyperbolic character \(e^{\rmh \myh
  t}=\cosh (\myh t)+\rmh\sinh(\myh t)\), see
\cite{Kisil10a}*{\eqref{E-eq:schroedinger-rep-conf-hyp}}, and obtain
the hyperbolic representation of \(\Space{H}{1}\),
cf.~\eqref{eq:schroedinger-rep-conf}:  
\begin{equation}
  \label{eq:schroedinger-rep-conf-hyp}
    [\uir{\rmh}{\myh}(s',x',y') \hat{f}\,](q)=e^{\rmh\myh (s'-x'y'/2)
    +\rmh x' q}\,\hat{f}(q-\myh y').  
\end{equation}
The corresponding derived representation is
\begin{equation}
  \label{eq:schroedinger-rep-conf-der-hyp}
  \uir{\rmh}{\myh}(X)=\rmh q,\qquad \uir{\rmh}{\myh}(Y)=-\myh \frac{d}{dq},
  \qquad
  \uir{\rmh}{\myh}(S)=\rmh\myh I.
\end{equation}
Then the associated Shale--Weil derived  representation of \(\algebra {sp}_2\) in
the Schwartz space \(\FSpace{S}{}(\Space{R}{})\) is, cf.~\eqref{eq:shale-weil-der}:
\begin{equation}
  \label{eq:shale-weil-der-double}
  \uir{\text{SW}}{\myh}(A) =-\frac{q}{2}\frac{d}{dq}-\frac{1}{4},\quad
  \uir{\text{SW}}{\myh}(B)=\frac{\rmh\myh}{4}\frac{d^2}{dq^2}-\frac{\rmh q^2}{4\myh},\quad
  \uir{\text{SW}}{\myh}(Z)=-\frac{\rmh\myh}{2}\frac{d^2}{dq^2}-\frac{\rmh q^2}{2\myh}.
\end{equation}
Note that \(\uir{\text{SW}}{\myh}(B)\) now generates a usual harmonic
oscillator, not the repulsive one like 
\(\uir{\text{SW}}{\myhbar}(B)\) in \eqref{eq:shale-weil-der}. 
However, the expressions in the quadratic algebra are still the same (up to a factor),
cf.~(\ref{eq:quadratic-A}--\ref{eq:quadratic-Z}):
\begin{eqnarray}
  \label{eq:quadratic-A-hyp}
  \qquad\uir{\text{SW}}{\myh}(A) &=&
  -\frac{\rmh}{2\myh}(\uir{\rmh}{\myh}(X)\uir{\rmh}{\myh}(Y)
  -{\textstyle\frac{1}{2}}\uir{\rmh}{\myh}(S))
  =-\frac{\rmh}{4\myh}(\uir{\rmh}{\myh}(X)\uir{\rmh}{\myh}(Y)
  +\uir{\rmh}{\myh}(Y)\uir{\rmh}{\myh}(X)),\\ 
  \label{eq:quadratic-B-hyp}
  \uir{\text{SW}}{\myh}(B) &=&
  \frac{\rmh}{4\myh}(\uir{\rmh}{\myh}(X)^2-\uir{\rmh}{\myh}(Y)^2), \\
  \label{eq:quadratic-Z-hyp}
  \uir{\text{SW}}{\myh}(Z)
  &=&-\frac{\rmh}{2\myh}(\uir{\rmh}{\myh}(X)^2+\uir{\rmh}{\myh}(Y)^2). 
\end{eqnarray}
This is due to the Principle~\ref{pr:similarity-correspondence} of
similarity and correspondence: we can swap operators \(Z\) and \(B\) with
simultaneous replacement of hypercomplex units \(\rmi\) and \(\rmh\).

The eigenspace of the operator \(2\uir{\text{SW}}{\myh}(B)\) with an
eigenvalue \(\rmh \nu\) are spanned by the Weber--Hermite
functions \(D_{-\nu-\frac{1}{2}}\left(\pm\sqrt{\frac{2}{\myh}}x\right)\),
see~\cite{ErdelyiMagnusII}*{\S~8.2}.  Functions \(D_\nu\) are
generalisations of the Hermit functions~\eqref{eq:hermit-poly}.

The compatibility condition for a ladder operator within the Lie algebra
\(\algebra{h}_1\) will be~\eqref{eq:hyp-ladder-compatib} as before,
since it depends only on the
commutators~(\ref{eq:cross-comm}--\ref{eq:cross-comm1}). Thus we still
have the set of ladder operators corresponding to values
\(\lambda=\pm1\):
\begin{displaymath}
  \ladder[h]{\pm}=\tilde{X}\mp\tilde{Y}=\rmh q\pm\myh \frac{d}{dq}.
\end{displaymath}
Admitting double numbers, we have an extra way to satisfy
\(\lambda^2=1\) in~\eqref{eq:hyp-ladder-compatib} with values
\(\lambda=\pm\rmh\).  Then there is an additional pair of hyperbolic
ladder operators, which are identical (up to factors)
to~\eqref{eq:ell-ladder-heisen-rep}:
\begin{displaymath}
  \ladder[\rmh]{\pm}=\tilde{X}\mp\rmh\tilde{Y}=\rmh q\pm\rmh\myh \frac{d}{dq}.
\end{displaymath}
Pairs \(\ladder[h]{\pm}\) and \(\ladder[\rmh]{\pm}\) shift
eigenvectors in the ``orthogonal'' directions changing their
eigenvalues by \(\pm1\) and \(\pm\rmh\).  Therefore an indecomposable
\(\algebra{sp}_2\)-module can be para\-metrised by a two-dimensional
lattice of eigenvalues in double numbers, see
Table~\ref{tab:2D-lattice}.
\begin{table}[htbp]
  \centering
\(  \xymatrix@R=2.5em@C=1.5em@M=.5em{
    & 
    \,\ldots\, \ar@<.4ex>[d]^{\ladder[\rmh]{+}} &  
    \,\ldots\, \ar@<.4ex>[d]^{\ladder[\rmh]{+}} & 
    \,\ldots\,  \ar@<.4ex>[d]^{\ladder[\rmh]{+}}  & 
    \\
    \ldots\, \ar@<.4ex>[r]^-{\ladder[h]{+}} & 
    \,V_{(n-1)+\rmh (k-1)}\,  \ar@<.4ex>[l]^-{\ladder[h]{-}}\ar@<.4ex>[r]^{\ladder[h]{+}}
     \ar@<.4ex>[u]^{\ladder[\rmh]{-}} \ar@<.4ex>[d]^{\ladder[\rmh]{+}} &  
    \,V_{n+\rmh (k-1)}\, \ar@<.4ex>[l]^{\ladder[h]{-}} \ar@<.4ex>[r]^{\ladder[h]{+}}
     \ar@<.4ex>[u]^{\ladder[\rmh]{-}} \ar@<.4ex>[d]^{\ladder[\rmh]{+}} & 
    \,V_{(n+1)+\rmh (k-1)}\,\ar@<.4ex>[l]^{\ladder[h]{-}}  \ar@<.4ex>[r]^-{\ladder[h]{+}}
     \ar@<.4ex>[u]^{\ladder[\rmh]{-}} \ar@<.4ex>[d]^{\ladder[\rmh]{+}}    & 
    \,\ldots\ar@<.4ex>[l]^-{\ladder[h]{-}}\\
    \ldots\, \ar@<.4ex>[r]^-{\ladder[h]{+}} & 
    \,V_{(n-1)+\rmh k}\,  \ar@<.4ex>[l]^-{\ladder[h]{-}}\ar@<.4ex>[r]^{\ladder[h]{+}}
     \ar@<.4ex>[u]^{\ladder[\rmh]{-}} \ar@<.4ex>[d]^{\ladder[\rmh]{+}} &  
    \,V_{n+\rmh k}\, \ar@<.4ex>[l]^{\ladder[h]{-}} \ar@<.4ex>[r]^{\ladder[h]{+}}
     \ar@<.4ex>[u]^{\ladder[\rmh]{-}} \ar@<.4ex>[d]^{\ladder[\rmh]{+}}& 
    \,V_{(n+1)+\rmh k}\,\ar@<.4ex>[l]^{\ladder[h]{-}}  \ar@<.4ex>[r]^-{\ladder[h]{+}}
     \ar@<.4ex>[u]^{\ladder[\rmh]{-}} \ar@<.4ex>[d]^{\ladder[\rmh]{+}}    & 
    \,\ldots\ar@<.4ex>[l]^-{\ladder[h]{-}}\\
    \ldots\, \ar@<.4ex>[r]^-{\ladder[h]{+}} & 
    \,V_{(n-1)+\rmh (k+1)}\,  \ar@<.4ex>[l]^-{\ladder[h]{-}}\ar@<.4ex>[r]^{\ladder[h]{+}}
     \ar@<.4ex>[u]^{\ladder[\rmh]{-}} \ar@<.4ex>[d]^{\ladder[\rmh]{+}} &  
    \,V_{n+\rmh (k+1)}\, \ar@<.4ex>[l]^{\ladder[h]{-}} \ar@<.4ex>[r]^{\ladder[h]{+}} 
     \ar@<.4ex>[u]^{\ladder[\rmh]{-}} \ar@<.4ex>[d]^{\ladder[\rmh]{+}}& 
    \,V_{(n+1)+\rmh (k+1)}\,\ar@<.4ex>[l]^{\ladder[h]{-}}  \ar@<.4ex>[r]^-{\ladder[h]{+}}
     \ar@<.4ex>[u]^{\ladder[\rmh]{-}} \ar@<.4ex>[d]^{\ladder[\rmh]{+}}    & 
    \,\ldots\ar@<.4ex>[l]^-{\ladder[h]{-}}\\
      & 
    \,\ldots\, \ar@<.4ex>[u]^{\ladder[\rmh]{-}} &  
    \,\ldots\, \ar@<.4ex>[u]^{\ladder[\rmh]{-}} & 
    \,\ldots\,  \ar@<.4ex>[u]^{\ladder[\rmh]{-}}  & }
\)
\caption{The action of hyperbolic ladder operators on a 2D
  lattice of eigenspaces. Operators \(\ladder[h]{\pm}\) move
  the eigenvalues by \(1\), 
  making shifts in the horizontal direction. Operators
  \(\ladder[\rmh]{\pm}\) change the eigenvalues by \(\rmh\), 
  shown as vertical shifts.}  
  \label{tab:2D-lattice}
\end{table}

The following functions 
\begin{eqnarray*}
  v_{\frac{1}{2}}^{\pm\myh}(q)&=&e^{\mp\rmh
    q^2/(2\myh)}=\cosh\frac{q^2}{2\myh}\mp \rmh\sinh \frac{q^2}{2\myh},\\
  v_{\frac{1}{2}}^{\pm\rmh}(q)&=&e^{\mp  q^2/(2\myh)}
\end{eqnarray*}
are null solutions to the operators \(\ladder[h]{\pm}\) and
\(\ladder[\rmh]{\pm}\), respectively. They are also eigenvectors of
\(2\uir{\text{SW}}{\myh}(B)\) with eigenvalues \(\mp\frac{\rmh}{2}\)
and \(\mp\frac{1}{2}\) respectively. If these functions are used as
mother wavelets for the wavelet transforms generated by the Heisenberg
group, then the image space will consist of the null-solutions of the
following differential operators,
see~\cite{Kisil10c}*{Cor.~\ref{C-co:cauchy-riemann}}:
\begin{displaymath}\textstyle
  D_{h}=\overline{X^{r} - Y^{r}}=(\partial_{ x} -\partial_{y})+\frac{\myh}{2}(x+y),
\qquad
  D_{\rmh}=\overline{X^{r} - \rmh Y^{r}}=(\partial_{ x} +\rmh\partial_{y})-\frac{\myh}{2}(x-\rmh
y),
\end{displaymath}
for \(v_{\frac{1}{2}}^{\pm\myh}\) and \(v_{\frac{1}{2}}^{\pm\rmh}\),
respectively. This is again in line with the classical
result~\eqref{eq:CR-Bargmann}. However annihilation of the eigenvector
by a ladder operator does not mean that the part of the 2D-lattice becomes
void, since it can be reached via alternative routes. Instead of
multiplication by a zero, as it happens in the elliptic case,
a half-plane of eigenvalues will be multiplied by the divisors of zero
\(1\pm\rmh\). 

We can also search ladder operators within the algebra
\(\algebra{sp}_2\) and admitting double numbers we will again find two sets
of them~\cite{Kisil09c}*{\S~\ref{W-sec:correspondence}}:
\begin{eqnarray*}
  \ladder[2h]{\pm} &=&\pm\tilde{A}+\tilde{Z}/2 =
   \mp\frac{q}{2}\frac{d}{dq}\mp\frac{1}{4}- \frac{\rmh\myh}{4}\frac{d^2}{dq^2}-\frac{\rmh q^2}{4\myh}=-\frac{\rmh}{4\myh}(\ladder[h]{\pm})^2,\\
  \ladder[2\rmh]{\pm}&=&\pm\rmh\tilde{A}+\tilde{Z}/2=  
  \mp\frac{\rmh q}{2}\frac{d}{dq}\mp\frac{\rmh}{4}-\frac{\rmh\myh}{4}\frac{d^2}{dq^2}-\frac{\rmh q^2}{4\myh}=-\frac{\rmh}{4\myh}(\ladder[\rmh]{\pm})^2.
\end{eqnarray*}
Again the operators \(\ladder[2h]{\pm}\) and \(\ladder[2h]{\pm}\) produce
double shifts in the orthogonal directions on the same two-dimensional
lattice in Tab.~\ref{tab:2D-lattice}.
    
\section{Ladder Operator for the Nilpotent Subgroup}
\label{sec:nilpotent-subgroup}

Finally, we look for ladder operators for the Hamiltonian
\(\tilde{B}+\tilde{Z}/2\) or, equivalently,
\(-\tilde{B}+\tilde{Z}/2\). It can be identified with a free
particle~\cite{Wulfman10a}*{\S~3.8}. 

We can look for ladder operators in the
representation~(\ref{eq:schroedinger-rep-conf-der}--\ref{eq:shale-weil-der})
within the Lie algebra \(\algebra{h}_1\) in the form
\(\ladder[\rmp]{\pm}=a\tilde{X}+b\tilde{Y}\). This is possible if and only if
\begin{equation}
  \label{eq:compatib-parab}
  -b=\lambda a,\quad 0=\lambda b.
\end{equation}
The compatibility condition \(\lambda^2=0\) implies \(\lambda=0\)
within complex numbers. However, such a ``ladder'' operator produces
only the zero shift on the eigenvectors, cf.~\eqref{eq:ladder-action}.

Another possibility appears if we consider the representation of the
Heisenberg group induced by dual-valued characters. On the
configurational space such a representation
is~\cite{Kisil10a}*{\eqref{E-eq:schroedinger-rep-conf-par}}: 
\begin{equation}
  \label{eq:schroedinger-rep-conf-par}
    [\uir{\rmp}{\chi}(s,x,y) f](q)= e^{2\pi\rmi x q}\left(
    \left(1-\rmp\myh (s-{\textstyle\frac{1}{2}}xy)\right) f(q)
    +\frac{\rmp\myh y}{2\pi\rmi} f'(q)\right).
\end{equation}
The corresponding derived representation of \(\algebra{h}_1\) is 
\begin{equation}
  \label{eq:schroedinger-rep-conf-der-par}
  \uir{p}{\myh}(X)=2\pi\rmi q,\qquad
  \uir{p}{\myh}(Y)=\frac{\rmp\myh}{2\pi \rmi} \frac{d}{dq},
  \qquad
  \uir{p}{\myh}(S)=-\rmp\myh I.
\end{equation}
However the Shale--Weil extension generated by this representation is
inconvenient.  It is better to consider the FSB--type parabolic
representation~\cite{Kisil10a}*{\eqref{E-eq:dual-repres}} on the phase
space induced by the same dual-valued character, cf.~\eqref{eq:stone-inf}:
\begin{equation}
  \label{eq:dual-repres}
  [\uir{\rmp}{\myh}(s,x,y)f](q,p)= e^{-2\pi\rmi(xq+yp)}(f(q,p)
  +\rmp\myh(s f(q,p) +\frac{y}{4\pi\rmi}f'_q(q,p)-\frac{x}{4\pi\rmi}f'_p(q,p))).
\end{equation}
Then the derived representation of \(\algebra{h}_1\) is:
\begin{equation}
  \label{eq:schroedinger-rep-conf-der-par1}
  \uir{p}{\myh}(X)=-2\pi\rmi q-\frac{\rmp\myh}{4\pi\rmi}\partial_{p},\qquad
  \uir{p}{\myh}(Y)=-2\pi\rmi p+\frac{\rmp\myh}{4\pi\rmi}\partial_{q},
  \qquad
  \uir{p}{\myh}(S)=\rmp\myh I.
\end{equation}
An advantage of the FSB representation is that the
derived form of the parabolic Shale--Weil representation coincides
with the elliptic one~\eqref{eq:shale-weil-der-ell}.

Eigenfunctions with the eigenvalue \(\mu\) of the parabolic
Hamiltonian \(\tilde{B}+\tilde{Z}/2=q\partial_p\) have the form
\begin{equation}
  \label{eq:par-eigenfunctions}
  v_\mu (q,p)=e^{\mu p/q} f(q), \text{ with an arbitrary function }f(q).
\end{equation}

The linear equations defining the corresponding ladder operator
\(\ladder[\rmp]{\pm}=a\tilde{X}+b\tilde{Y}\) in the algebra
\(\algebra{h}_1\) are~\eqref{eq:compatib-parab}.  The compatibility
condition \(\lambda^2=0\) implies \(\lambda=0\) within complex numbers
again. Admitting dual numbers, we have additional values
\(\lambda=\pm\rmp\lambda_1\) with \(\lambda_1\in\Space{C}{}\) with the
corresponding ladder operators
\begin{displaymath}
  \ladder[\rmp]{\pm}=\tilde{X}\mp\rmp\lambda_1\tilde{Y}=
  -2\pi\rmi q-\frac{\rmp\myh}{4\pi\rmi}\partial_{p}\pm 2\pi\rmp\lambda_1\rmi p= 
  -2\pi\rmi q+   \rmp\rmi( \pm 2\pi\lambda_1 p+\frac{\myh}{4\pi}\partial_{p}).
\end{displaymath}
For the eigenvalue \(\mu=\mu_0+\rmp\mu_1\) with \(\mu_0\),
\(\mu_1\in\Space{C}{}\) the
eigenfunction~\eqref{eq:par-eigenfunctions} can be rewritten as:
\begin{equation}
  \label{eq:par-eigenfunctions-1}
  v_\mu (q,p)=e^{\mu  p/q} f(q)= e^{\mu_0  p/q}\left(1+\rmp\mu_1
    \frac{p}{q}\right) f(q)
\end{equation}
due to the nilpotency of \(\rmp\).  Then the ladder action of
\(\ladder[\rmp]{\pm}\) is \(\mu_0+\rmp\mu_1\mapsto \mu_0+\rmp(\mu_1\pm
\lambda_1)\).  Therefore, these operators are suitable for building
\(\algebra{sp}_2\)-modules with a one-dimensional chain of
eigenvalues.

Finally, consider the ladder operator for the same element \(B+Z/2\)
within the Lie algebra \(\algebra{sp}_2\). According to the above
procedure we get the equations:
\begin{displaymath}
  -b+2c=\lambda a,\qquad
  a=\lambda b,\qquad
  \frac{a}{2}=\lambda c,
\end{displaymath}
which can again be resolved if and only if \(\lambda^2=0\). There is
the only complex root \(\lambda=0\) with the corresponding operators
\(\ladder[p]{\pm}=\tilde{B}+\tilde{Z}/2\), which does not affect
the eigenvalues.  However the dual number roots \(\lambda
=\pm\rmp\lambda_2\) with \(\lambda_2\in\Space{C}{}\) lead
to the operators
\begin{displaymath}
  \ladder[\rmp]{\pm}=\pm \rmp\lambda_2\tilde{A}+\tilde{B}+\tilde{Z}/2
  = \pm\frac{\rmp\lambda_2}{2}\left(q\partial_{q}-p\partial_{p}\right)+q\partial_{p}. 
\end{displaymath}

\section{Conclusions: Similarity and Correspondence}
\label{sec:concl-simil-corr}

We wish to summarise our findings. Firstly, the appearance of
hypercomplex numbers in ladder operators for \(\algebra{h}_1\) follows
exactly the same pattern as was already noted for
\(\algebra{sp}_2\)~\cite{Kisil09c}*{Rem.~\ref{W-re:hyper-number-necessity}}:
\begin{itemize}
\item the introduction of complex numbers is a necessity for the
  \emph{existence} of ladder operators in the elliptic
  case;
\item in the parabolic case, we need dual numbers to make
  ladder operators \emph{useful};
\item in the hyperbolic case, double numbers are not required 
  neither for the existence or for the usability of ladder operators, but
  they do provide an enhancement. 
\end{itemize}
In the spirit of the Similarity and Correspondence
Principle~\ref{pr:similarity-correspondence} we have the following
extension of Prop.~\ref{W-pr:ladder-sim-eq} from~\cite{Kisil09c}: 
\begin{prop}
  \label{pr:ladder-sim-eq}
  Let a vector \(H\in\algebra{sp}_2\) generates the subgroup \(K\),
  \(N'\) or \(\Aprime\), that is \(H=Z\), \(B+Z/2\), or
  \(2B\), respectively. Let \(\alli\) be the respective hypercomplex
  unit. Then the ladder operators \(\ladder[]{\pm}\)  satisfying to the
    commutation relation:
  \begin{displaymath}
    [H,\ladder[2]{\pm}]=\pm\alli \ladder{\pm}%,\qquad [\ladder{-},\ladder{+}]=2\alli H.
  \end{displaymath}
  are given by:
  \begin{enumerate}
  \item Within the Lie algebra \(\algebra{h}_1\): \(\ladder{\pm}=\tilde{X}\mp\alli \tilde{Y}.\)
  \item Within the Lie algebra \(\algebra{sp}_2\): \(
    \ladder[2]{\pm}=\pm\alli \tilde{A} +\tilde{E}\).
  Here \(E\in\algebra{sp}_2\) is a linear combination of  \(B\) and
  \(Z\) with the properties:
  \begin{itemize}
  \item \(E=[A,H]\).
  \item \(H=[A,E]\).
  \item Killings form \(K(H,E)\)~\cite{Kirillov76}*{\S~6.2} vanishes.
  \end{itemize}
  Any of the above properties defines the vector \(E\in\loglike{span}\{B,Z\}\)
  up to a real constant factor.
  \end{enumerate}
\end{prop}
It is worth continuing this investigation and describing in detail
hyperbolic and parabolic versions of FSB spaces.

\textbf{Acknowledgements:} I am grateful to the anonymous referees for
their helpful remarks. 

\small
\bibliography{abbrevmr,akisil,analyse,algebra,arare,aclifford,aphysics}
\end{document}